\documentclass[twocolumn]{jpsj2} 
\usepackage[dvips]{color}

\title{Polarized Superfluidity in the imbalanced attractive Hubbard model}

\author{Akihisa \textsc{Koga}$^{1}$
\thanks{E-mail address: koga@phys.titech.ac.jp}
and Philipp \textsc{Werner}$^{2}$}

\inst{$^{1}$Department of Physics, Tokyo Institute of Technology, Tokyo 152-8551, Japan \\
$^{2}$Theoretische Physik, 
ETH Zurich, Z\"urich 8093, Switzerland}

\abst{
We investigate the attractive Hubbard model in infinite spatial dimensions 
by means of dynamical mean-field theory. 
Using a continuous-time Monte Carlo algorithm in the Nambu formalism
as an impurity solver, we directly deal with 
the superfluid phase in the population imbalanced system. 
By calculating the superfluid order parameter, the magnetization, 
and the density of states, we discuss 
how the polarized superfluid state is realized
in the attractive Hubbard model at quarter filling.
We find that a drastic change in the density of states is induced by 
spin imbalanced populations in the superfluid state.
}

\kword{Superfluid, imbalanced system, continuous-time Monte Carlo simulation}

\begin{document}
\maketitle
\section{Introduction}
The superfluid state in ultracold atomic systems
has attracted much interest since the successful realization of
the Bose-Einstein condensation (BEC) of rubidium atoms.\cite{Rb} 
In addition to bosonic systems, 
the superfluid state has been observed in two-component fermionic systems,
\cite{Regal}
where Cooper pairs formed 
by the attractive interactions condense at low temperatures.
Due to the high controllability of the interaction strength and 
the particle number, 
interesting phenomena have been observed such as 
the BCS-BEC crossover\cite{BCSBEC1,BCSBEC2,BCSBEC3} and 
the superfluid state with imbalanced populations.
\cite{Imbalance1,Imbalance2}
These observations stimulate further experimental and theoretical investigations
on fermionic systems.

In the existing literature on spin imbalanced populations 
various ordered ground states have been proposed to be more stable than 
the polarized superfluid (PSF) state,
which is naively expected to be realized below the critical temperature.
One interesting candidate is the Fulde-Ferrell-Larkin-Ovchinnikov (FFLO) 
phase,\cite{FFLO1,FFLO2}
where Cooper pairs are formed with nonzero total momentum.
This phase has been observed in the high field region 
in ${\rm CsCoIn_5},$\cite{CsCoIn1,CsCoIn2,CsCoIn3}
and has theoretically been discussed in the latter compound,\cite{FFLOH} 
as well as cold atoms with imbalanced populations.\cite{FFLOO,FFLO1D}
Another proposed phase is 
the breached-pair (BP) phase, where
both the superfluid order parameter and the magnetization are finite 
at zero temperature. \cite{BP1,BP2,BP3,BP4,BP5}
When one considers higher dimensional optical lattice systems,
the BP state without momentum dependence may be one of 
the appropriate ground states.
It has recently been clarified that 
the PSF state is closely 
connected to the BP phase at half filling 
in the three-dimensional Hubbard model 
with intermediate attractive interactions. \cite{ImbalanceDao}
However, the Hubbard model has a high symmetry at half filling, 
\cite{TD1,TD2,Freericks}
and the conclusions may not be applicable to an optical lattice system,
where the particle density is not fixed at half filling due to the existence of 
the confining potential.
Therefore, it is important to clarify how the PSF state and
the BP state are realized in a system away from half filling.

With this purpose in mind, we investigate the attractive Hubbard model
at quarter filling to discuss the effect of 
the imbalanced spin populations on the superfluid state. 
By combining dynamical mean-field theory (DMFT)
\cite{Metzner,Muller,Georges,Pruschke}
with the continuous time quantum Monte Carlo (CTQMC) method,\cite{Rubtsov}
we study the low temperature properties of the system quantitatively.
Here, we extend the CTQMC method in the continuous-time auxiliary field (CTAUX) formulation\cite{CTAUX}
to treat the PSF state in the Nambu formalism.
By calculating the order parameter of the superfluid state, the magnetization,
and the density of states, we clarify the nature of the PSF state
in the spin imbalanced system.

The paper is organized as follows.
In \S2, we introduce the model Hamiltonian for the attractive Hubbard model and 
briefly summarize the DMFT framework. 
The CTQMC algorithm in the Nambu formalism is
explained in some detail in \S3. 
In \S4, we focus on the attractive Hubbard model at quarter filling 
to discuss how the PSF state is realized at low temperatures.
A brief summary is given in \S5.

\section{Model and Method}

We consider a correlated fermion system with attractive interactions,
which may be described by the Hubbard Hamiltonian,
\begin{equation}
\hat{\cal{H}}=\sum_{(i,j),\sigma}
\left[-t-\left(\mu+h\sigma\right)\delta_{ij}\right]c^{\dagger}_{i\sigma}c_{j\sigma}
-U\sum_{i}n_{i\uparrow}n_{i\downarrow},
\label{eq1}
\end{equation}
where $c_{i\sigma}$ ($c^{\dagger}_{i\sigma}$) 
is an annihilation (creation) operator of a fermion on the $i$th site
with spin $\sigma$, and 
$n_{i\sigma}= c^{\dagger}_{i\sigma}c_{i\sigma}$. 
$U$ is the onsite attractive interaction, $t$ is the transfer integral
between sites, $\mu$ is the chemical potential, and $h$ is the magnetic field.
For $h=0$ the ground state properties of the model have been studied 
in one dimension,\cite{Lieb,Shiba,Machida,Xianlong,Pour,Fujihara} two dimensions\cite{TD1,TD2,Koga} 
and infinite dimensions.\cite{Garg,Toschi,Suzuki,Keller,Bauer,Freericks}
Both the BCS-BEC crossover and 
the possibility of a supersolid state
have been discussed.
On the other hand, there are few studies addressing the effect of 
imbalanced populations beyond the static mean-field approach 
except for one dimensional system.\cite{FFLO1D}

To study the infinite dimensional attractive Hubbard model 
at an arbitrary filling, we make use of DMFT.
\cite{Metzner,Muller,Georges,Pruschke}
In DMFT, the original lattice model is mapped to an effective impurity model, 
which accurately takes into account local particle correlations.
The lattice Green's function is obtained via a self-consistency condition imposed
on the impurity problem.
This treatment is formally exact in infinite dimensions, and 
the DMFT method has successfully been applied to strongly correlated fermion systems.

When the superfluid state is treated in the framework of DMFT, 
the local self-energy should be described by a $2\times 2$ matrix as
\begin{equation}
\hat{\Sigma}(i\omega_n)=\left(
\begin{array}{cc}
\Sigma_\uparrow(i\omega_n)&S(i\omega_n)\\
S(i\omega_n)&-\Sigma^*_\downarrow(i\omega_n)
\end{array}
\right),
\end{equation}
where $\Sigma_\sigma(i\omega_n)\; [S(i\omega_n)]$ is 
the diagonal (off-diagonal) 
element of the self-energy in the Nambu formalism and 
the Matsubara frequency is $\omega_n = (2n+1)\pi/\beta$, with  
$\beta$ the inverse temperature.
Note that
we do not take into account $k$-dependent correlations,
but dynamical correlations
through the frequency-dependent self-energy.
This enables us to 
discuss the stability of the $s$-wave superfluid state more quantitatively
beyond the static mean-field theory.

The lattice Green's function is then given in terms of the local self-energy as,
\begin{eqnarray}
\hat{G}^{-1}(k, i\omega_n)=\left(i\omega_n+h\right)\hat{\sigma}_0+
\left(\mu-\epsilon_k\right)\hat{\sigma}_z-\hat{\Sigma}\left(i\omega_n\right),
\end{eqnarray}
where $\hat{\sigma}_0$ and $\hat{\sigma}_z$ are the identity matrix and 
the $z$-component of the Pauli matrix, and 
$\epsilon_k$ is the dispersion relation for the non-interacting
system.
The local lattice Green's function is obtained as,
\begin{eqnarray}
\hat{G}(i\omega_n) = \int dk \hat{G}(k, i\omega_n).
\end{eqnarray}
In the calculations, we use the semi-circular density of states, 
$\rho(x) = 1/N\sum_k\delta (x-\epsilon_k) = 2/\pi D \sqrt{1-(x/D)^2}$, 
where $D$ is the half bandwidth. 
The self-consistency equation\cite{GeorgesZ} is then given by
\begin{equation}
\hat{G}_{0,\text{imp}}^{-1}(i\omega_n) = \left(i\omega_n +h \right)\hat{\sigma}_0
+\mu\hat{\sigma}_z-\left(\frac{D}{4}\right)^2\hat{\sigma}_z\hat{G}(i\omega_n)\hat{\sigma}_z\label{eq:self}.
\end{equation}

When one discusses low energy properties in strongly correlated systems
in the framework of DMFT, an impurity solver is necessary 
to obtain the Green's function
and the self-energy for the effective impurity model.
There are various numerical techniques such as 
exact diagonalization\cite{Caffarel} and
the numerical renormalization group.\cite{NRG,NRG_RMP,OSakai}
A recently developed and particularly powerful method is CTQMC.
In this method, Monte Carlo samplings of collections of diagrams for the partition function
are performed in continuous time,
and thereby the Trotter error, which originates from the Suzuki-Trotter
decomposition, is avoided. 
Furthermore, this method is applicable to more general classes of models than the Hirsch-Fye algorithm.\cite{Hirsch}
The CTQMC method has successfully been applied to 
various systems such as the Hubbard model,\cite{CTQMC,Multi} 
the periodic Anderson model, \cite{Luitz}
the Kondo lattice model\cite{Otsuki}
and the Holstein-Hubbard model.\cite{Phonon}

\section{Continuous-Time Quantum Monte Carlo simulations in the Nambu Formalism}

In this section, we explain the CTAUX method,\cite{CTAUX} and extend it  
to treat the superfluid state.
A similar solver was recently proposed,~\cite{Luitz}
where the superfluid state is treated by means of a canonical transformation.
The Anderson impurity model we have to solve is given by
\begin{eqnarray}
H&=&H_0 + H_U,\label{eq:anderson}\\
H_0&=&\sum_{p\sigma} \epsilon_{p\sigma} n_{p\sigma}
+\sum_{p\sigma}\left(V_{p\sigma} d_\sigma^\dag a_{p\sigma}+h.c.\right)
\nonumber\\
&+&\sum_{p}\left(\Delta_p a_{p\uparrow}^\dag a_{p\downarrow}^\dag +h.c.\right)
+\sum_\sigma E_{d\sigma} n_{d\sigma},\\
H_U&=&-U\left[n_{d\uparrow} n_{d\downarrow}
-\frac{1}{2}\left(n_{d\uparrow}+n_{d\downarrow}-1\right)\right],
\end{eqnarray}
where $a_{p\sigma} (d_\sigma)$ annihilates a fermion with spin $\sigma$ 
in the $p$th orbital of the effective baths (the impurity site).
$\epsilon_{p\sigma}$ and $\Delta_p$ represent the effective bath, and 
$V_{p\sigma}$ represents the hybridization 
between the effective bath and the impurity site. 
$E_{d\sigma}$ is the energy level for the impurity site,
$n_{p\sigma}=a_{p\sigma}^\dag a_{p\sigma}$, and
$n_{d\sigma}=d_\sigma^\dag d_\sigma$. 
We note that the total particle number is not conserved 
in the model.
The Green's functions should be defined by 
$\hat{G}(\tau)=\langle T_\tau \hat{\psi}(\tau) \hat{\psi}^\dag(0)\rangle$, 
where $T_\tau$ is the imaginary-time ordering operator and
$\hat{\psi}^\dag(\tau)=(c^\dag_\uparrow(\tau)\; c_\downarrow(\tau) )$. 
The Green's functions are $2\times 2$ matrices with elements 
\begin{equation}
\hat{G}(\tau)=\left(
\begin{array}{cc}
G_\uparrow(\tau)&F(\tau)\\
F^*(\tau)&-G_\downarrow(-\tau)
\end{array}
\right),
\end{equation}
where $G_\sigma(\tau) = \langle T_\tau c_\sigma(\tau)
c_\sigma^\dag(0)\rangle$ denotes the normal Green's function, 
and  
$F(\tau) = \langle T_\tau c_\uparrow(\tau) c_\downarrow(0)\rangle$ and
$F^*(\tau) = \langle T_\tau c^\dag_\downarrow(\tau)
c^\dag_\uparrow(0)\rangle$ anomalous Green's functions.
Here, we have chosen the Green's functions $G_\sigma(\tau)$ to be positive.

To perform simulations, 
we consider here a weak coupling CTQMC approach.
The partition function $Z$ is given by
\begin{eqnarray}
Z&=&{\rm Tr}\left[e^{-\beta H_1} T_\tau e^{-\int_0^\beta d\tau H_2(\tau)} 
\right]\nonumber\\
&=& \sum_{n=0}^\infty \int_0^\beta d\tau_1 \int_{\tau_1}^\beta d\tau_2 \cdots \int_{\tau_{n-1}}^\beta d\tau_n \nonumber\\
&\times&{\rm Tr}\Big[
e^{-(\beta-\tau_n)H_1}(-H_2)e^{-(\tau_n-\tau_{n-1})H_1}\cdots\nonumber\\
&&\cdots
e^{-(\tau_2-\tau_1)H_1}(-H_2)e^{-\tau_1H_1}
\Big],
\end{eqnarray}
where we have divided the impurity Hamiltonian Eq.~(\ref{eq:anderson})
into two parts as,
\begin{eqnarray}
H_1&=&H-H_2,\\
H_2&=&H_U-K/\beta\nonumber\\
&=&\frac{K}{2\beta}\sum_{s=-1,1} e^{\gamma s \left(n_\uparrow+n_\downarrow-1\right)},
\end{eqnarray}
with $\gamma= \cosh^{-1} (1+\beta U/2 K)$, and $K$ some nonzero constant.
The introduction of the Ising variable $s$ in $H_2$ 
enables us to perform simulations away from half-filling. 
An $n$th order configuration
$c = \{ s_1, s_2, \cdots , s_n; \tau_1, \tau_2, \cdots, \tau_n\}$
corresponding to auxiliary spins $s_1,s_2,\ldots,s_n$ at imaginary times $\tau_1<\tau_2<\ldots<\tau_n$
contributes a weight
\begin{eqnarray}
w_c &=& e^{-K}\left(\frac{Kd\tau}{2\beta}\right)^n
e^{-\gamma \sum s_i}Z_0\;{\rm det}\; 
\left[\hat{N}^{(n)}\right]^{-1}
\end{eqnarray}
to the partition function.
Here, $Z_0={\rm Tr}[e^{-\beta H_1}]$ and $\hat{N}^{(n)}$ is an $n\times n$ matrix, 
where each element consists of a $2\times 2$ matrix:
\begin{eqnarray}
\left[\hat{N}^{(n)}\right]^{-1} &=& \hat{\Gamma}^{(n)}
-\hat{g}^{(n)}\left(\hat{\Gamma}^{(n)}-\hat{I}^{(n)}\right),\\
\hat{I}^{(n)}_{ij}&=&\delta_{ij}\hat{\sigma}_0,\\ 
\hat{\Gamma}^{(n)}_{ij} &=& \delta_{ij} e^{\gamma s_i}\hat{\sigma}_0,\\
\hat{g}^{(n)}_{ij} &=&\left(
\begin{array}{cc}
g_{0\uparrow}(\tau_i-\tau_j)&f_0(\tau_i-\tau_j)\\
-f_0^*(\tau_i-\tau_j)&g_{0\downarrow}(\tau_j-\tau_i)
\end{array}
\right),\hspace{4mm}
\end{eqnarray}
with $i,j = 1,2, \cdots n$.

The sampling process must satisfy
ergodicity and (as a sufficient condition) detailed balance.
For ergodicity, it is enough to insert or remove the Ising variables
with random orientations at random times to generate all possible configurations.
To satisfy the detailed balance condition, we decompose the transition probability as
\begin{eqnarray}
p\left(i\rightarrow j\right) = p^{\rm prop}\left(i\rightarrow
j\right)p^{\rm acc}\left(i\rightarrow j\right),
\end{eqnarray}
where $p^{\rm prop} (p^{\rm acc})$ is the probability to propose (accept) 
the transition from the configuration $i$ to the configuration $j$. 
Here, we consider the insertion and removal of the Ising spins 
as one step of the simulation process, 
which corresponds to a change of $\pm 1$ in the perturbation order.
The probability of insertion/removal of an Ising spin is then given by 
\begin{eqnarray}
p^{\rm prop}(n\rightarrow n+1)&=&\frac{d\tau}{2\beta},\\
p^{\rm prop}(n+1\rightarrow n)&=&\frac{1}{n+1}.
\end{eqnarray}
For this choice, the ratio of the acceptance probabilities becomes
\begin{eqnarray}
\frac{p^{\rm acc}\left(n\rightarrow n+1\right)}
{p^{\rm acc}\left(n+1\rightarrow n\right)}
= \frac{K}{n+1}e^{-\gamma s_{n+1}}\frac{\det N^{(n)}}
{\det N^{(n+1)}}.
\end{eqnarray}
When the Metropolis algorithm is used to sample the configurations, 
we accept the transition 
from $n$ to $n\pm1$ with the probability
\begin{eqnarray}
{\rm min}\left[1, \frac{p^{\rm acc}\left(n\rightarrow n\pm 1\right)}
{p^{\rm acc}\left(n\pm 1\rightarrow n\right)}\right].
\end{eqnarray}


In each Monte Carlo step, we measure the following Green's functions ($0<\tau<\beta$),
\begin{eqnarray}
G_\sigma(\tau)&=&\frac{1}{Z}{\rm Tr}
\left[e^{-\beta H}c_\sigma(\tau)c^\dag_\sigma(0)\right],\\
F(\tau)&=&\frac{1}{Z}{\rm Tr}
\left[e^{-\beta H}c_\uparrow(\tau)c_\downarrow(0)\right],\\
F^*(\tau)&=&\frac{1}{Z}{\rm Tr}
\left[e^{-\beta H}c^\dag_\downarrow(\tau)c^\dag_\uparrow(0)\right].
\end{eqnarray}
By using Wick's theorem, the contribution of a certain configuration $c$ is given by 
\begin{align}
G_{\sigma}^{c}(\tau) &= \det [N^{(n)}] 
{\rm det}\left(
\begin{array}{cc}
\left[N^{(n)}\right]^{-1}&Q_\sigma\\
R_\sigma&g_{0\sigma}(\tau)
\end{array}
\right),\\
F^{c}(\tau) &= \det [N^{(n)}] 
{\rm det}\left(
\begin{array}{cc}
\left[N^{(n)}\right]^{-1}&Q'\\
R'& f_0(\tau)
\end{array}
\right),\\
F^{*c}(\tau) &= \det [N^{(n)}] 
{\rm det}\left(
\begin{array}{cc}
\left[N^{(n)}\right]^{-1}&Q^{*\prime}\\
R^{*\prime}& f^{*}_0(\tau)
\end{array}
\right),
\end{align}
where $Q_\sigma, Q^{\prime}, Q^{*\prime}, R_\sigma, R^\prime, R^{*\prime}$ are vectors,
in which the $i$th element ($i=1,2,\cdots, n$) is defined by
\begin{eqnarray}
Q_{\uparrow i} &=& \{-g_{0\uparrow}(\tau_i)\;\; f^*_0(\tau_i)\}^T,\\
Q_{\downarrow i} &=& \{f_0(\tau_i-\tau)\;\; g_{0\downarrow}(\tau-\tau_i)\}^T,\\
Q'_i &=& \{-f_0(\tau_i)\;\; -g_{0\downarrow}(-\tau_i)\}^T,\\
Q^{*\prime}_i &=& Q_{\uparrow i},\\
R_{\uparrow i}&=& (e^{\gamma s_i}-1)\{g_{0\uparrow}(\tau-\tau_i)\;\; 
f_0(\tau-\tau_i)\},\\
R_{\downarrow i}&=& (e^{\gamma s_i}-1)\{f^*_0(-\tau_i)\;\;
-g_{0\downarrow}(\tau_i)\},\\
R'_i &=& R_{\uparrow i},\\
R^{*\prime}_i &=& (e^{\gamma s_i}-1)\{f^*_0(\tau-\tau_i)\;\;
-g_{0\downarrow}(\tau_i-\tau)\}.\hspace{2mm}
\end{eqnarray}
In this paper, we use the half bandwidth $D$ as the unit of the energy
and set $K=1$ in the CTQMC simulations.
We thus calculate static physical quantities such as 
the order parameter of the superfluid state $\Delta$ and 
the magnetization $m$, which are defined by
\begin{eqnarray}
\Delta&=&\langle c_\uparrow c_\downarrow \rangle=F(0_+),\\
m&=&\sum_\sigma \sigma\langle c_\sigma^\dag c_\sigma \rangle = -\sum_\sigma \sigma G_\sigma(0_+).
\end{eqnarray}
Furthermore, by applying the maximum entropy method (MEM) 
to the Green's functions, 
we deduce the spectral functions, which allows us to discuss 
static and dynamical properties of the system.

\begin{figure}[htb]
\begin{center}
\includegraphics[width=7cm]{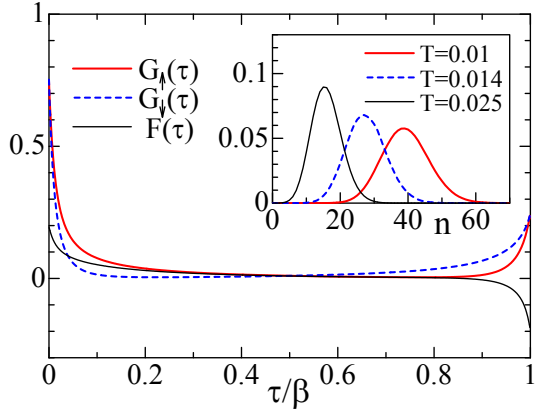}
\caption{Green's functions as a function of $\tau/\beta$
in the quarter-filled system at $U=1, h=0.1$ and $T=0.01$.
The solid (dashed) line represents the Green's function for the up (down) spin and
the thin line the anomalous Green's function. 
The inset shows the probability distribution
for configurations with perturbation order $n$ 
at the temperatures $T=0.01, 0.014$ and $0.025$.
}
\label{fig:Green}
\end{center}
\end{figure}

In Fig. \ref{fig:Green}, we show, as an example, 
the normal and anomalous Green's
functions when $U=1$, $h=0.1$ and $T=0.01$. 
The Green's functions were measured on a grid of a thousand points.
In this case, the system has both a magnetization 
$m\sim 0.005$ and 
a superfluid order parameter $\Delta\sim 0.2$.
Therefore, we can say that the PSF state is realized 
in this parameter region.
Note that a large difference appears between 
$G_\uparrow(\tau)$ and $G_\downarrow(\tau)$ 
near $\tau \sim 0$ and $\beta$ although the magnetization is small.
This may affect dynamical properties.

\section{Superfluid state in a magnetic field}
\begin{figure}[b]
\begin{center}
\includegraphics[width=7cm]{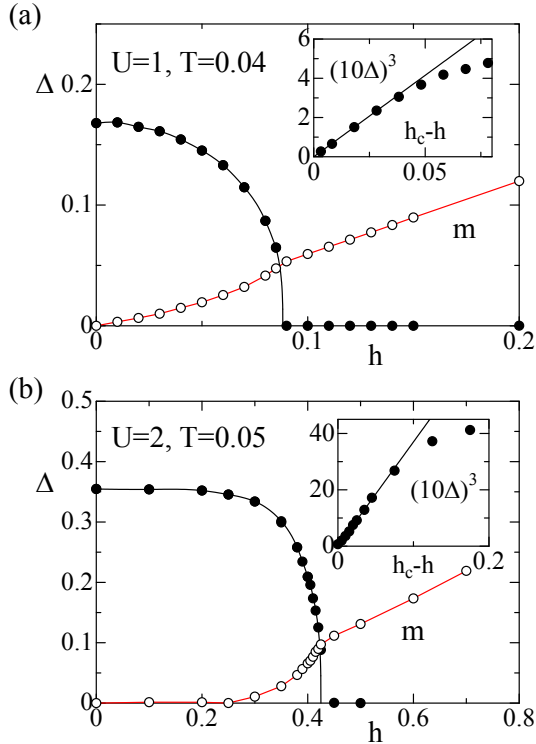}
\caption{The superfluid order parameter and the magnetization as a
function of the magnetic field when $U=1, T=0.04$ (a), and 
$U=2, T=0.05$ (b).
The insets show the critical behavior for the order parameter.
}
\label{fig:fig3}
\end{center}
\end{figure}

\begin{figure}[htb]
\begin{center}
\includegraphics[width=7cm]{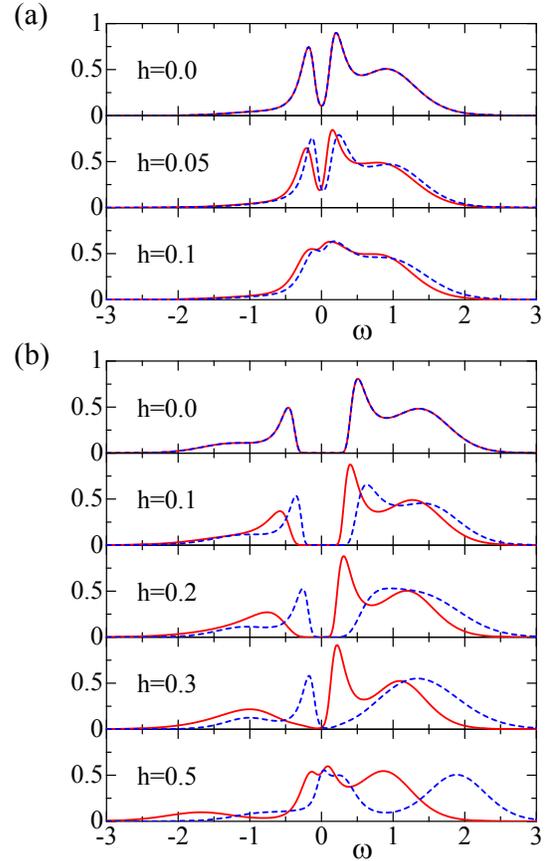}
\caption{Solid (dashed) lines represent the spectral functions for fermions 
with up (down) spin when $U=1, T=0.04$ (a), and $U=2, T=0.05$ (b).
}
\label{fig:dos}
\end{center}
\end{figure}

Here, we focus on the attractive Hubbard model at quarter filling
to discuss how the PSF state is realized at low temperatures.
First, we perform calculations at a fixed temperature. 
Results for the systems with weak (intermediate) coupling
$[U=1~(U=2)]$ are shown in Fig. \ref{fig:fig3}.
When no magnetic field is applied,
the system is in the superfluid state at low temperatures.
In fact, we find that 
the superfluid gap opens around the Fermi level
and that peak structures appear at the edges of the gap
in the density of states, as shown in Fig. \ref{fig:dos}.
These results are consistent with those 
obtained by other groups.\cite{Garg,Bauer} 
If a magnetic field is applied to the system, 
these peaks move to low (high) energy region 
in the density of states for up (down) spin.
Pairing correlations are then suppressed, and a magnetization is induced, as shown in Fig. \ref{fig:fig3}.
We note that at low temperatures,
the introduction of a magnetic field has little effect on the 
static quantities $\Delta$ and $m$, but produces a drastic change 
in the density of states.
In fact, it is found that when $U=2$ and $h=0.3$, 
one of the peaks disappears and the other remains above (below) the Fermi level 
in the density of states for up (down) spin
although the superfluid gap is still open.
Therefore, we conclude that 
dynamical properties are strongly affected by the spin imbalanced populations.
A further increase in the magnetic field smears
the superfluid gap around the Fermi level and 
the superfluid order parameter vanishes.
This suggests the existence of a phase transition to the normal metallic phase.
By examining the critical behavior $\Delta \sim |h-h_c|^{1/\delta}$
with the exponent $\delta = 3$, 
we obtain the critical fields $h_c(U=1, T=0.04) \sim 0.0885$ and 
$h_c(U=2, T=0.05) \sim 0.425$, as shown in the insets of Fig. \ref{fig:fig3}.
It is also found that
the phase transition induces a cusp singularity 
in the magnetization curve.
The results obtained here are in contrast to those in the half-filled attractive 
Hubbard model on the simple cubic lattice, 
where the PSF state smoothly 
connects to the normal metallic phase.\cite{ImbalanceDao}
This may result from the fact that 
the competition between the superfluid state and 
the charge density wave state 
enhances fluctuations for the superfluid order parameter 
due to the high symmetry at half filling.
It would be interesting to clarify this point, 
which is beyond the scope of our study.

\begin{figure}[t]
\begin{center}
\includegraphics[width=7cm]{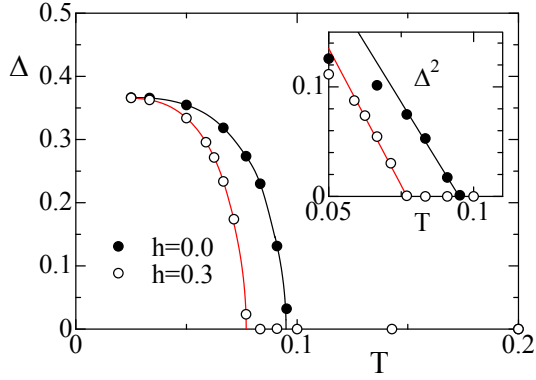}
\caption{The order parameter for the superfluid state $\Delta$ as a
function of the temperature $T$ when $U=2$. 
}
\label{fig:fig1}
\end{center}
\end{figure}

We also show the temperature dependence of the superfluid order parameter 
in Fig. \ref{fig:fig1}.
When $h=0$, as temperature is decreased, 
the order parameter $\Delta$ appears where 
the phase transition occurs from the normal metallic state 
to the superfluid state. 
By examining the critical behavior $\Delta \sim |T-T_c|^\beta$ with the exponent
$\beta=1/2$, 
we obtain the critical temperature $T_c\sim 0.095$,
as shown in the inset of Fig. \ref{fig:fig1}.
On the other hand, when the magnetic field is switched on,
pairing correlations are suppressed.
For $h=0.3$, it is found that 
the superfluid order parameter is decreased and 
the critical temperature is shifted to $T_c \sim 0.077$.
A large magnetic field destroys the superfluidity and 
the normal metallic state is realized instead.
In fact, we could not find any finite $\Delta$
down to low temperatures ($T=0.033$) in the case $h=0.6$.
Note that the two curves saturate almost at the same value of $\Delta$ 
when $T\rightarrow 0$. 
This suggests that by increasing the magnetic field at zero temperature, 
the superfluid ground state is little affected and eventually a 
first order phase transition occurs to the normal metallic state.
Therefore, it may be difficult to realize the BP state with 
finite magnetization.

\begin{figure}[b]
\begin{center}
\includegraphics[width=7cm]{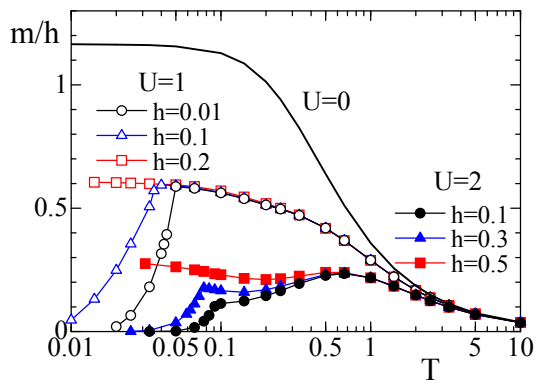}
\caption{Normalized magnetization $m/h$ as a
function of temperature $T$ in the system with $U=0, 1$ and $2$. 
}
\label{fig:fig2}
\end{center}
\end{figure}

\begin{figure}[t]
\begin{center}
\includegraphics[width=7cm]{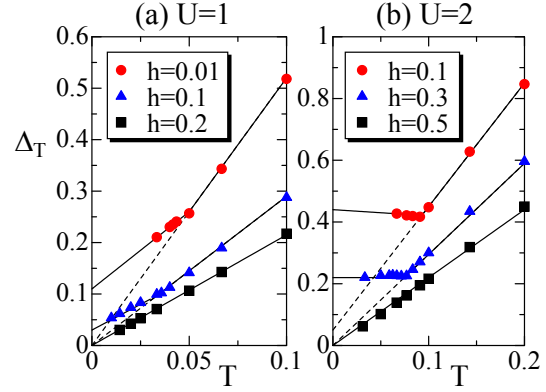}
\caption{$\Delta_T (=-T\log m)$ as a
function of temperature $T$ in the system with $U=1$ (a) and $U=2$ (b). 
Lines are guides to eyes.
}
\label{fig:exp}
\end{center}
\end{figure}

To clarify this, 
we next examine how the magnetization appears in the superfluid state.
In Fig. \ref{fig:fig2}, we show a semi-log plot of the magnetization 
normalized by the applied field.
When a tiny magnetic field is applied to the system, 
$m/h$ corresponds to the magnetic susceptibility $\chi$.
In the noninteracting system ($U=0$), 
$m/h$ saturates at low temperatures at the value 
$\chi(T=0) = 2 \rho(x =0)\sim 1.16$, and in the interacting case,
our results are consistent with 
those obtained by Keller {\it et al}.\cite{Keller}
Increasing the attractive interaction in the presence of a finite magnetic field,
fermion pairs are formed at high temperatures and thereby
magnetic correlations are suppressed and the magnetization decreases.
When the magnetic field is small enough,
a phase transition occurs to the superfluid state at low temperatures.
In this state,
the magnetization rapidly decreases below the critical temperature, 
as shown in Fig. \ref{fig:fig2}.
This means that it is difficult to realize the BP ground state with
finite $\Delta$ and $m$ at zero temperature.
To confirm this, 
we also show the quantity $\Delta_T(=-T \log m)$ as a function 
of temperature in Fig. \ref{fig:exp}.
When $T\rightarrow 0$,
the data approach a finite value in the superfluid state, while
they approach zero in the normal metallic state.
This means that
the magnetization decays exponentially in $1/T$ in the superfluid state.
Therefore, we can say that 
the BP state is not realized in the ground state,
at least, in this quarter-filled system.

\begin{figure}[b]
\begin{center}
\includegraphics[width=7cm]{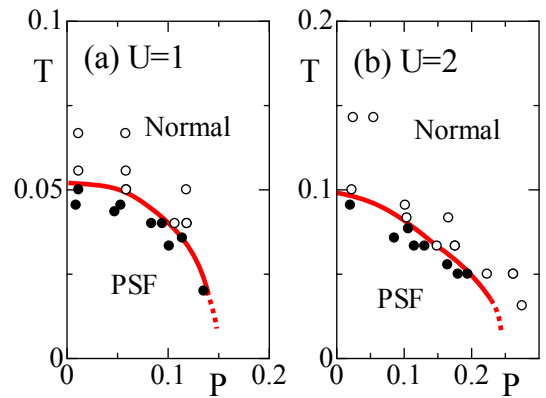}
\caption{Phase diagrams for the quarter-filled system with $U=1$ and $U=2$. 
Open (solid) circles indicate the normal (PSF) state and 
phase boundaries are guides to eyes.
}
\label{fig:phase}
\end{center}
\end{figure}

\begin{figure}[htb]
\begin{center}
\includegraphics[width=7cm]{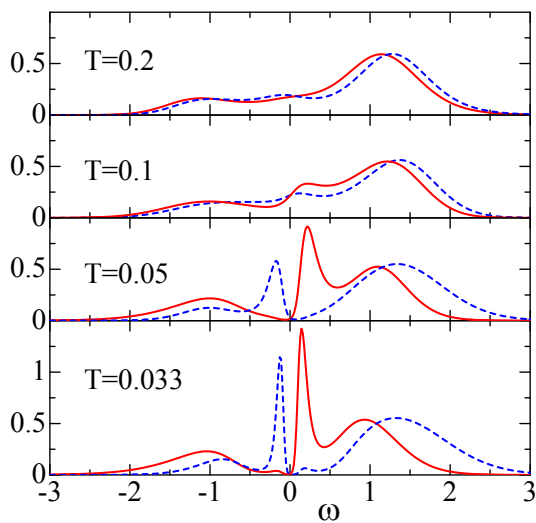}
\caption{Density of states for the system with $U=2$ and $P=0.02$
when $T=0.2, 0.1, 0.05$ and $0.033$.
}
\label{fig:dos2}
\end{center}
\end{figure}

By performing similar calculations, we have obtained the phase diagram
for the spin imbalance parameter $P [=(N_\uparrow-N_\downarrow)/(N_\uparrow+N_\downarrow) = 2m]$, 
which is sometimes used in the discussion of optical lattice systems, 
as shown in Fig. \ref{fig:phase}.
When the temperature decreases with fixed imbalanced populations,
a phase transition occurs to the PSF state.
Figure~\ref{fig:dos2} shows the density of states for each spin component 
in a system with $P\sim 0.02$ and $U=2$.
It is found that at high temperatures $(T>T_c)$,
the normal metallic state is realized, where the spin imbalanced populations
have little effect on the density of states.
By contrast, in the superfluid state $(T<T_c)$,
a large difference appears
in the low energy region of the spectral functions.
Since the populations with up spin are slightly larger than 
those with down spin in the case considered, 
a certain energy is necessary to add a fermion with up spin 
in the superfluid state.
Therefore the low energy peak appears above the Fermi level 
in the density of states for up spin at low temperatures. 
As temperature is lowered to zero, it may be difficult to realize a state 
with $P\neq 0$, as discussed before. 
We cannot rule out that a phase transition from the superfluid phase 
back to the metallic phase (reentrant behavior) will occur 
at temperatures below the range accessible to us. 
In any event, the imbalanced populations should play a crucial role 
at very low temperatures, in particular, in the dynamical properties.

\section{Summary}
We have investigated the attractive Hubbard model in infinite dimensions
by means of DMFT. 
Here, we have used the CTAUX method as an impurity solver, 
which has been extended to treat the superfluid state directly 
in the Nambu formalism.
We have calculated the superfluid order parameter, the magnetization, and 
the density of states systematically 
to discuss how the PSF state is realized at low temperatures.
It was found that when the temperature is lowered 
in the presence of a fixed magnetic field, 
a superfluid phase transition indeed occurs in our model,
and the magnetization exponentially decays in the superfluid state. 
This suggests that the BP phase is unstable at zero temperature.
We have also found that a drastic change in the density of states is induced by 
spin imbalanced populations in the superfluid state although 
the spin imbalance has little effect on static quantities.
It is an interesting problem to clarify how 
such dynamical properties are realized 
in a low dimensional optical lattice with a confining potential,
which is now under consideration.

\section*{Acknowledgment}
The authors thank J. Bauer, N. Kawakami, K. Okunishi, and Th. Pruschke 
for valuable discussions. 
Parts of the computations were done on TSUBAME
Grid Cluster at the Global Scientific Information and Computing
Center of the Tokyo Institute of Technology. 
This work was partly supported by the Grant-in-Aid for Scientific Research 
20740194 (A.K.) and 
the Global COE Program ``Nanoscience and Quantum Physics" from 
the Ministry of Education, Culture, Sports, Science and Technology (MEXT) 
of Japan. PW acknowledges support from SNF Grant PP002-118866.

\end{document}